# Space-Time Video Regularity and Visual Fidelity: Compression, Resolution and Frame Rate Adaptation

Dae Yeol Lee, Hyunsuk Ko, Jongho Kim, and Alan C. Bovik, Fellow, IEEE

*Abstract*—In order to be able to deliver today's voluminous amount of video contents through limited bandwidth channels in a perceptually optimal way, it is important to consider perceptual trade-offs of compression and space-time downsampling protocols. In this direction, we have studied and developed new models of natural video statistics (NVS), which are useful because high-quality videos contain statistical regularities that are disturbed by distortions. Specifically, we model the statistics of divisively normalized difference between neighboring frames that are relatively displaced. In an extensive empirical study, we found that those paths of space-time displaced frame differences that provide maximal regularity against our NVS model generally align best with motion trajectories. Motivated by this, we build a new video quality prediction engine that extracts NVS features from displaced frame differences, and combines them in a learned regressor that can accurately predict perceptual quality. As a stringent test of the new model, we apply it to the difficult problem of predicting the quality of videos subjected not only to compression, but also to downsampling in space and/or time. We show that the new quality model achieves state-of-the-art (SOTA) prediction performance compared on the new ETRI-LIVE Space-Time Subsampled Video Quality (STSVQ) database, which is dedicated to this problem. Downsampling protocols are of high interest to the streaming video industry, given rapid increases in frame resolutions and frame rates.

## I. INTRODUCTION

THE media industry is steadily improving the realism of video experiences streamed to the consumers by expanding the ranges of video space along all dimensions. Consumer video contents are being acquired and streamed at increasingly higher spatial resolutions, frame rates, and dynamic ranges (HDR). Media streaming services like Amazon Prime Video, Netflix, and YouTube now deliver high-quality 4K/60fps/HDR television and cinematic content to consumer, and high-motion content, such as sports, is causing content providers to consider even higher frame rates. Display manufacturers are ahead of the game, and televisions and monitors that support 8K HDR and

D. Lee, and A.C. Bovik are with the Department of Electrical and Computer Engineering, The University of Texas at Austin, Austin, TX, USA (email:daelee711@utexas.edu, bovik@ece.utexas.edu). H. Ko is with the Division of Electrical Engineering, Hanyang University ERICA, Ansan, South Korea (email: hyunsuk@hanyang.ac.kr). J. Kim is with the Realistic AV Research Group, ETRI, Daejeon, South Korea (email: pooney@etri.re.kr).

true 120Hz video signal playout are available, albeit currently expensive. Indeed, recent high-end smartphones and tablets have bright displays supporting HDR and refresh rates of 120Hz. It is natural to expect that these cycles of increments of video dimensions and launches of sharper, faster, and deeper displays that support them will continue, towards meeting the seemingly insatiable demand for more realistic, immersive, high performance media delivery.

Increases of video dimensionality inevitably leads to enormous data volumes, presenting significant challenges to content providers seeking to deliver them over limited bandwidth channels in a perceptually satisfactory way. The principal technology enabling bandwidth-constrained delivery is video compression, as exemplified by the ITU standards H.264 [1], HEVC [2], and VVC [3], and open-source standards like VP9 [4] and AV1 [5]. While video compression technologies effectively reduce the data volumes, they also introduce annoying compression artifacts, especially in a limited bit budget environment [6]. A second enabling technology are globally deployed perceptual video quality prediction like SSIM [7] and VMAF [8], which are used to balance the perception-bandwidth tradeoff. Nevertheless, as video data volumes and streaming popularity continue to explode, more creative compression augmentation protocols are needed. One recent approach currently being deployed by content providers is to combine video compression with spatial resolution adaptation, whereby spatial subsampling is applied before compression on some frames. Following decompression at the playback side, these frames are spatially upsampled before display. Subsampling decisions are typically made under the control of perceptual quality algorithms.

In this direction, various authors have studied this perceptual trade-off. In [9]-[11], the authors investigated the combined effects of spatial subsampling and compression on the perceptual quality of videos. More recently, the idea of also attempting temporal subsampling before compression has also been considered, given ongoing and future increases of frame rates. The authors of [12], [13] considered the effects of temporal subsampling on perceptual video quality, and proposed methods of frame rate adaptation. The authors of [14] proposed a space-time resolution adaptation method for video compression, but the decision to subsample was considered separately in space and time. However, while these studies have deepened our understanding of how spatial and temporal subsampling each individually affect perceptual quality, when used as precursors to compression, less works has been applied towards modelling the perceptual effects of simultaneously applying spatial and temporal subsampling protocols prior to applying compression.

Fortunately, very recent psychometric resources [15], [16] have become available that may advance our understanding of the joint perceptual effects, and tradeoffs, of spatial and temporal subsampling and compression. The AVT-VQDB-UHD-1 database [15] provides subjective opinion scores on 120 videos distorted by joint application of spatial and temporal subsampling and compression on 5 source contents. However, the maximum frame rate considered was 60Hz. The much larger ETRI-LIVE STSVQ database [16] provides a rich collection of contemporaneous resources, including subjective quality scores rendered on 4K 10-bit videos at frame rates up to 120Hz, processed by a wide range of levels of simultaneous spatial and temporal subsampling and compression using HEVC. The database contains scores on 437 space-time subsampled and compressed videos generated from 15 source contents.

In order to be able to conduct perceptually optimized rate control, what is needed are predictive models, that can be translated into practical algorithms, of the perceptual effects of combined compression, spatial, and temporal downsampling. Towards advancing progress in this direction, we propose a new video quality model based on our findings on the space-time statistics of videos. The new video quality model is able to account for varying degrees of spatial and temporal subsampling applied jointly with compression. The new model attains state-of-the-art (SOTA) performance on the new human study database. The contributions that we make are as follows.

- We present a new model of the space-time statistics of motion pictures. More specifically, we model the statistics of the differences of neighboring frames that are relatively displaced in space and time. Such displacements relate to motion, but also to visual information-gathering via small (microsaccadic) eye movements. Perceptual models that we deploy derive from temporal lag filtering in visual area of the thalamus LGN, and space-time contrast normalization in cortical area V1. We have discovered that space-time normalized differences possess a very high degree of inherent statistical regularity when displaced along the motion trajectory.
- We devised a way to identify space-time displacement paths that yield maximum statistical regularities, deploy new models of how these regularities are disturbed by distortions arising from, for example, subsampling in space and/or time, and/or compression. Using these we construct an entirely unique full reference (FR) video quality predictor of perceived space-time video distortions.
- The new video quality model is ideally suited to assist the emerging problem of conjoint space-time resolution adaptation strategies as a way of further optimizing perceptual streaming video compression.

The rest of the paper is organized as follows: In Section II, we discuss prior work on perceptual video quality prediction. In Section III, we describe our recent findings on the natural statistics of space-time displaced frame differences. In Section IV, we give a detailed description of our video quality model. In Section V, we compare and analyze the performances of the new model against relevant high-performance video quality models. Finally, we draw conclusions in Section VI.

## II. Related work

Being able to accurately measure perceptual quality has become recognized as an essential ingredient when designing and optimizing streaming media services. Over the years, a wide variety of objective image and video quality prediction models have been developed that target these needs. Video quality prediction models can be broadly classified as either Reference (including full reference and reduced reference) and No-reference models. The former assumes there is available complete or partial information derived from an existing reference pristine video, while the latter assumes that no such reference information can be accessed. While both classes are valuable tools for media quality optimizations, the control of video codecs is a dominant application. As such, here we focus on Reference models and applications, especially towards problems arising in the context of spatial and/or temporal dimension reduction (subsampling) methods applied in concert with compression.

The field of Reference video quality prediction includes such older frame-based (spatial) like MSE and PSNR, Structural SIMilarity index (SSIM) [7], and multi-scale SSIM (MSSSIM) [17], Visual Information Fidelity (VIF) index [18], Detail Loss Measure (DLM) [19], and Additive Impairment Measure (AIM) [19]. All of these mentioned models are widely used by the streaming video industry.

Frame based models like these can be applied to conduct video quality prediction aggregating frame predictions using some kind of temporal pooling [20]. However, aggregating spatial (frame) scores does not capture temporal video distortions. To remedy this, a variety of video quality models have been devised that use temporal features. The Video Quality Metric (VQM) [21], and variable frame delay sensitive version (VQM-VFD) [22], partition videos into small, short-duration space-time volumes, then extract simple features, such as spatial gradients and frame-differences, which are pooled to produce video quality predictions. ST-MAD [23] and ViS3 [24] measure motion artifacts on space-time slices of the original and distorted video volumes, comparing them using the Most Apparent Distortion (MAD) model [25]. ST-RRED [26] and SpEED [27] deploy natural video statistics models of statistical regularities inherent in video frames and frame-differences, and how they are altered by distortions. Video Multi-method Assessment Fusion (VMAF) [8] is a popular high-performance video quality model that fuses quality-aware video features from frame-differences, VIF, and DLM, using a Support Vector Regressor (SVR). These video quality models are able to achieve high prediction performances on popular subject video quality databases such as LIVE VQA [28], CSIQ [24], LIVE

Mobile VQA [29], and VQEG HD3 [30].

More recent quality studies have expanded their scopes to include the combined effect of video compression and spatial subsampling [9]-[11]. These studies did not consider another potential way of enhancing streaming video compression: temporal downsampling or frame rate reduction. Some work has been directed towards analyzing the quality of videos containing frame rate variations. The authors of [31] constructed the BVI-HFR video quality database, and used it to develop the Frame Rate dependent video Quality Metric (FRQM) [32], which predicts the effects of frame rate variations on perceptual quality, but without considering compression, nor spatial subsampling. The authors of [33] conducted a perceptual study on the combined effects of compression and frame rate variations and constructed the LIVE-YT-HFR video quality database. However, no study to date has considered the combined perception of compression, spatial subsampling, and temporal subsampling, a gap we aim to fill.

### III. On the Space-Time Statistics of Motion Pictures

A foundational concept of visual neuroscience is that the statistical properties of the visual environment have impacted the way our visual system transforms, encodes, and extracts information from visual signals [34]. Widely accepted models of natural scene statistics involve linear band-pass decompositions, which accounts for processes of scale and/or orientation sensitive decorrelation, followed by divisive normalization mechanisms, which approximate non-linear gain control in neurons along the visual pathway [35]-[40]. These transforms reveal an inherent statistical regularity of natural pictures and videos. The shapes of the distributions of the transformed signals strongly tends towards a Gaussian characteristic, in the absence of distortion.

Statistical models of natural videos have been used with great success in video quality prediction applications, where perceptual quality is inferred by quantifying distortion-induced deviations from these models [26], [27], [41]-[43]. Natural video statistics (NVS) models of both frames and frame differences have been used to capture spatial and temporal aspects of perceptual. NVS models only exist for frame differences without space-time directionality, which may fail to capture many aspects of space-time distortions and perception of them. Here, we broaden and deepen the modeling of frame difference statistics by introducing displacements in both space and time prior to differencing neighboring frames. We show that there exist strong space-time direction-dependent statistical regularities in motion pictures which we utilize to better model video quality.

#### A. Space-time Displaced Frame Differences

One reason why we are interested in space-time directional statistics that characterize displaced frame differences is that videos contain significant redundancies in the direction of local motions arising from projected object or camera movements. Another reason is that they may correlate with small, microsaccadic eye movements that occur around points of gaze, and which are theorized to help achieve more efficient visual encoding in the brain [44], [45]. Retinal signals are commonly modeled as being subjected to temporal lag filtering in Lateral Geniculate Nucleus (LGN), which is a form of smoothed temporal differencing operation [46]. Another good reason is that we have been able to show that very strong statistical regularities occur along motion field trajectories, and that these may be measured using tools derived from studies of natural video statistics [47].

To begin building our model, let luminance of the video frames be denoted as $I$. Given a space-time displacement vector $d = (x, y, t)$, spatially and temporally displaced frame differences between frames $k$ and $k + t$ may be generally expressed

$$I_{fd}(i, j, k) = I(i, j, k) - I(i + x, j + y, k + t), \quad (1)$$

where $(i, j, k)$ are constrained by the finite dimension of the video according to $i \in [\max(1, 1 - x), \min(W, W - x)]$, $j \in [\max(1, 1 - y), \min(H, H - y)]$, and $k \in [1, T - t]$, where $H$, $W$, and $T$ refer to the height, width, and number of frames of the video (or video clip or scene, as the case may be), respectively.

#### B. Divisive Normalization

The space-time displaced frame differences are subjected to divisive normalization, corresponding to non-linear gain control. In our model, the coefficients are computed as

$$\hat{I}_{fd}(i, j, k) = \frac{I_{fd}(i, j, k)}{\sigma(i, j, k) + C}, \quad (2)$$

where $I_{fd}$ are the displaced frame differences, $\sigma$ is the local weighted rms contrast, and $C$ is a saturation constant. Note that local contrast $\sigma$ is

$$\sigma(i, j, k) = \sqrt{\sum_{l=-L}^{L} \sum_{m=-M}^{M} \omega_{lm} \left[ I_{fd}(i + l, j + m, k) - \mu(i, j, k) \right]^2}, \quad (3)$$

where $\omega_{lm}$ is a symmetric Gaussian window sampled out to three standard deviations ($L=5$, $M=5$) and rescaled to unit volume, and where the weighted mean luminance functions $\mu$ is computed as

$$\mu(i, j, k) = \sum_{l=-L}^{L} \sum_{m=-M}^{M} \omega_{lm} I_{fd}(i + l, j + m, k). \quad (4)$$

#### C. Statistical Regularity Map Construction

It is perhaps not obvious at first that the shapes of the empirical distributions of the divisively normalized coefficients (1) heavily depend on the displacement vector $d = (x, y, t)$, or that proper choices of $d$ yields highly predictable, regular distributions while other choices do not. As it turns out, these direction-dependent statistical regularities tend to align with the directions of motion. As a first step towards demonstrating, and subsequently exploiting this property, we have developed a way to construct a statistical regularity map for finding a space-time path having maximum regularity.

As shown in Fig. 1, first partition each video frame into patches of size $N \times N$, and compute displaced frame differences using a range of displacement vectors constrained to $[-R, R]^2$, defined relative to the patch dimension $N$ and the temporal separation $t$: $R = (\lfloor N/6 \rfloor - \lfloor N/6 \rfloor \bmod 2) \times t$. The use of a limited search range reduces the computational complexity and is supported by the fact that per-frame velocity is generally limited to small magnitudes [48]-[50]. Of course, larger displacements may be considered. These displaced frame differences are subjected to divisive energy normalization, followed by scaling to unit variance. We have found that differences between frames displaced along the direction of motion strongly tend towards Gaussianity to a remarkable degree [47], but along other directions, they do not. Thus, empirical probability distribution of the coefficients obtained via the aforementioned processing steps are then compared agaisnt the canonical gaussian distribution $(\sim \mathcal{N}(0,1))$. We deploy the Kullback Leibler Divergence (KLD) to compare the distributions (empirical against ideal Gaussian model)

$$D_{KL}(P||Q) = \sum_i P(i) \log\left(\frac{P(i)}{Q(i)}\right), \quad (5)$$

where $P(i)$ and $Q(i)$ are the empirical probability densities of the transformed coefficients and the canonical gaussian, respectively. Each displacement location yields a corresponding KLD value, which forms a space-time "regularity map". The optimal vector that yields the maximal degree of regularity (Gaussianity) is determined by averaging those displacement vectors that yield the lowest 5$^{TH}$ percentile of KLD values on the space-time regularity map. This vector is deemed to correspond to the displacement direction most aligned with the local motion of the video, in the absence of distortion.

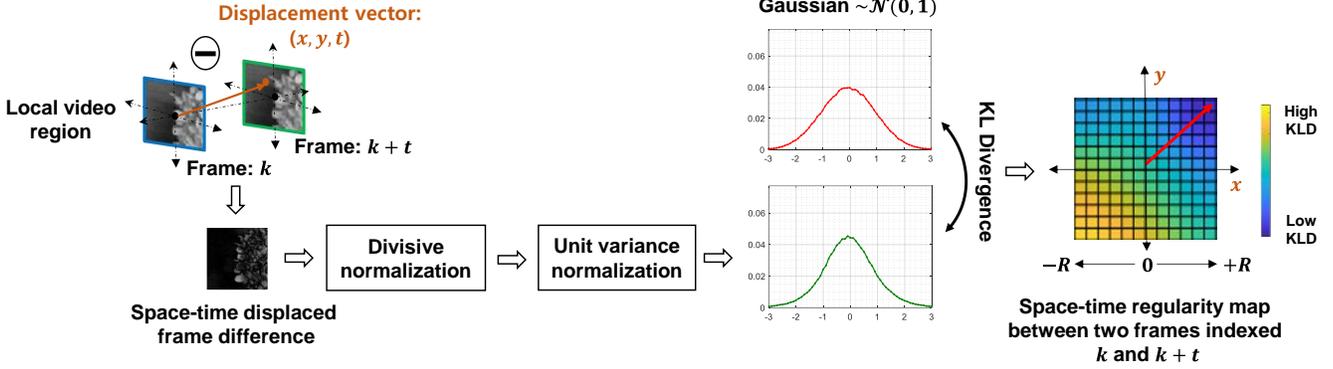

Fig. 1. Procedure for constructing a space-time regularity map. The displacement direction having the lowest KL divergence indicates the path associated with the most regular frame differences.

Fig. 2 illustrates examples of the computed "most regular" space-time paths of local space-time regions of videos from the Middlebury optical flow database [51], as well as the average ground truth motion vectors for each video patches. The space-time regularity map of Fig. 2(a) was constructed using a temporal separation $t = 1$, i.e., by differencing patches from adjacent frames that are spatially displaced. As shown in Fig. 2(a), the optimal spatial displacement that delivered maximum Gaussianity among the displaced frame difference coefficients was (0,7), in agreement with the average ground-truth motion direction of the video patch in Fig. 2(b). The case of a temporal separation $t = 2$ is depicted in Figs. 2(c) and 2(d). The average per-frame ground truth motion vector for this particular patch is (-3, -2), which can be denoted as $(-3t, -2t, t)$ for a frame

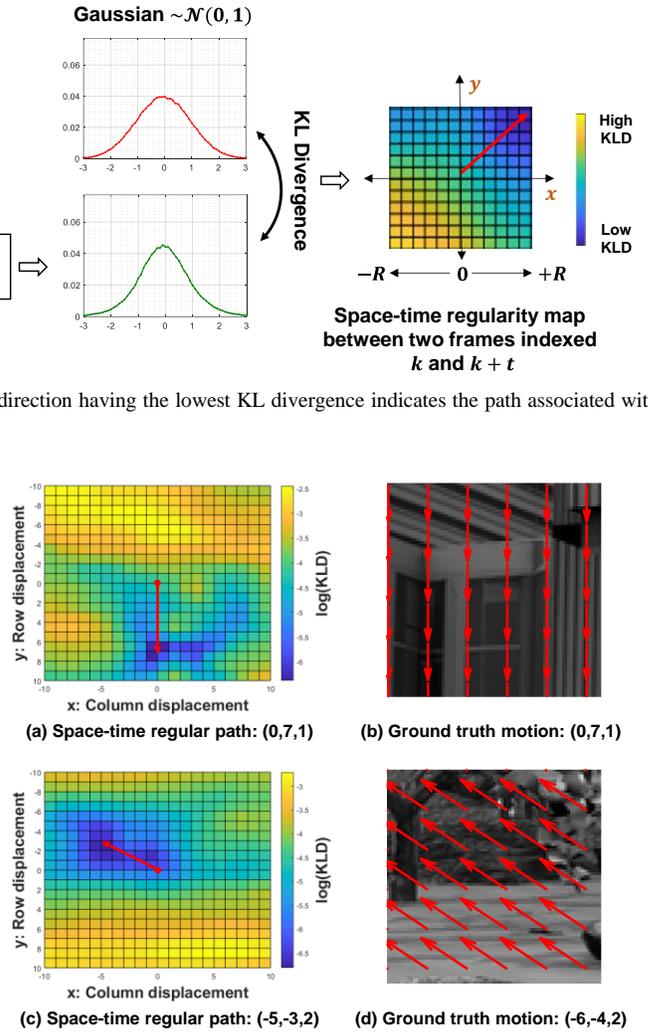

(a) Space-time regular path: (0,7,1)    (b) Ground truth motion: (0,7,1)

(c) Space-time regular path: (-5,-3,2)    (d) Ground truth motion: (-6,-4,2)

Fig. 2. (a), (c) Space-time regularity maps constructed on local space-time regions of videos from the Middlebury optical flow database. The optimal space-time regular path is indicated by a red arrow, with the vector values given below each figure. (b), (d) Visualization of a local video region with average ground truth motion vector indicated by red arrows.

separation $t$. It follows that the average ground truth motion vector of this patch for $t = 2$ is (-6, -4, 2), as shown in Fig. 2(d). The space-time regularity map constructed on this region for patches separated by two frames is shown in Fig. 2(c). It may be seen that the displacement path yielding the highest degree of regularity was (-5, -3, 2), which again well aligns with the true motion direction.

We further verified the same statistical regularities along the motion direction by observing the statistics of frame difference patches collected along many space-time displacement trajectories [47]. We used the HD1K optical flow database [52], which provides ground-truth motion vectors along multiple frames (~ 1 second duration) of 2560×1080 videos. Fig. 3 shows examples of video patches traced along various space-time displacement trajectories. Video patches of size 100×100 were collected for 41 frames, along three Types of trajectories: Type 1, "motion;" Type 2, "non-displaced;" and Type 3, "random drift." Type 1 trajectories trace the ground-truth motion; Type 2 trajectories maintain the same (initial) spatial coordinate throughout; and Type 3 trajectories consist of uniformly random displacements drawn from the 2D interval $[-20, 20]^2$. at each progression of frame. As a result, frame patch volumes of dimension 100×100×41 are formed for each Type of trajectory. Then, for each Type, adjacent frame patches are differenced, forming frame difference volumes which are then subjected to divisive normalization and scaling to unit variance. The distributions of the computed coefficients for each of the three Types of frame difference are plotted as red empirical probability density function (PDF) curves in Fig.3. Overlaid in blue are plots of $\mathcal{N}(0,1)$ densities. The KLD values between the computed coefficients and the ideal Gaussian model are provided above each distribution plots. For each example in Fig. 3, it may be observed that the coefficients collected along the motion trajectory adhered best to the Gaussian model.

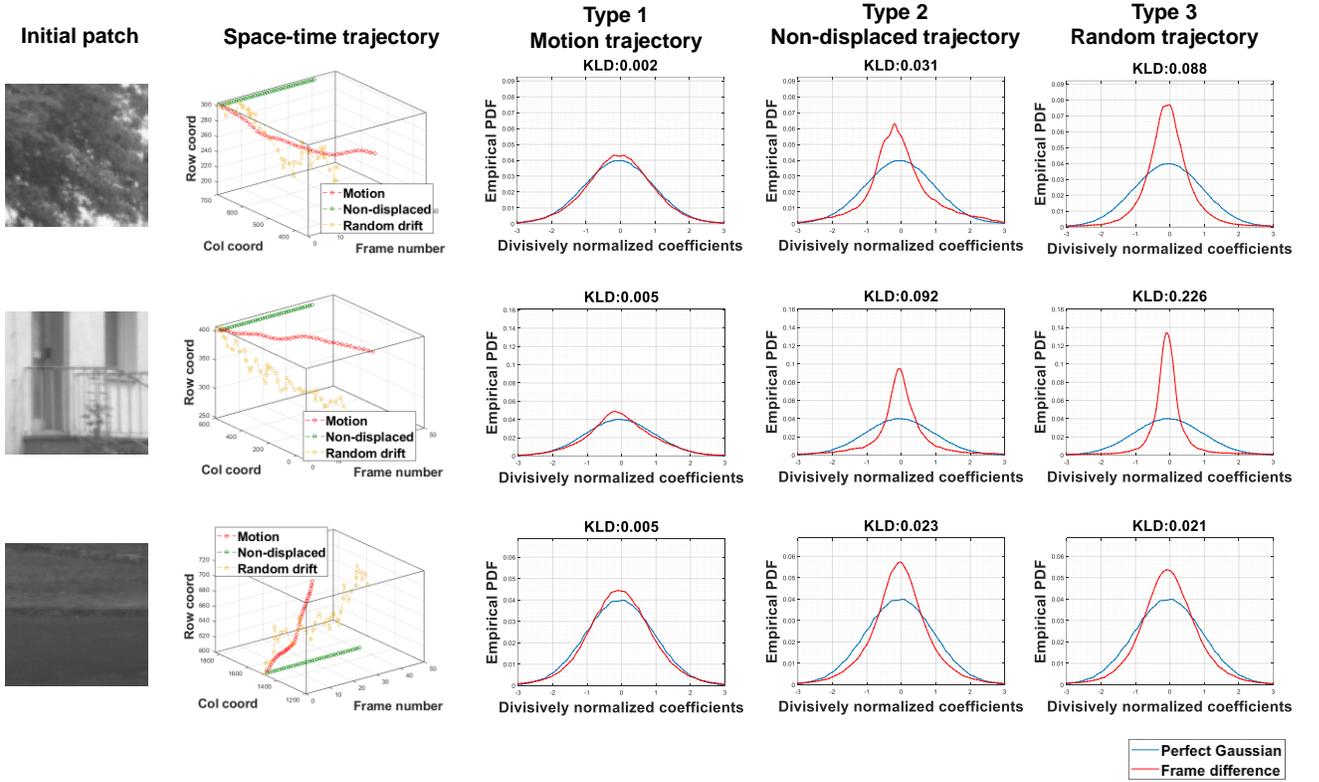

Fig. 3. Comparison of distributions between the canonical Gaussian and frame-differences displaced along various trajectories (Type 1, motion; Type 2, non-displaced; and Type3, random), followed by divisive normalization.

We have found that there exist space-time displacement paths of frame differences that reveal strong regularities that may be exploited. Indeed, this may be viewed as a basis for the success of motion-compensated video coding.

## IV. VIDEO QUALITY MODEL

Now we introduce a new video quality model we have developed that is based on statistical measurements space-time regularities. As such, we refer to it as the Video Space-Time Regularity (VSTR) model. An overall flowchart of VSTR is presented in Fig. 4. The proposed model first determines the "most regular" displacement vector, from the space-time regularity map of the reference video, thereby avoiding the effects of distortion. We then compare the space-time statistics of the test videos against those of the reference along the paths defined by the displacement vectors, to assess whether, and by how much, they have been disturbed by distortion. As explained in the foregoing, a set of quality-aware features that quantify the degree of statistical divergence between the space-time bandpass coefficients of the two videos are extracted, and combined using an SVR that is trained to predict the video quality of the distorted video.

### A. Displacement Vector Determination

First, we describe the determination of the displacement vectors in detail. Each optimal displacement vector is computed on every one-second segment of the reference video. By analyzing the initial portion (200msec) of each one second segment, we derive a dominant per-frame displacement vector at specified spatial patch coordinates that best reveals the "regularity" of differences computed between adjacent frames, using the criteria just described. This per-frame displacement vector also is later used to determine space-time displacement vectors for different amounts of temporal (frame) separation.

From the initial 200msec segment, we first select $N$ frames (we use $N$=3) temporally separated by 200msec/$N$. Then, we also collect the next frame after each selected frame, forming $N$ adjacent frame pairs. Since videos generally contain local motions presenting in many directions, we partition each collected frame into $M \times M$ sized patches (here, $M$ =301). Given all of the adjacent pairs of patches from consecutive frames, construct a space-time regularity map following the procedure detailed in Section III-C. Each frame patch pairs yields a displacement vector maximizing the regularity of the frame difference coefficients. This procedure resembles the concept of block-wise motion estimation; however, it has a different and specific aim: finding space-time paths having a type of optimal statistical regularity.

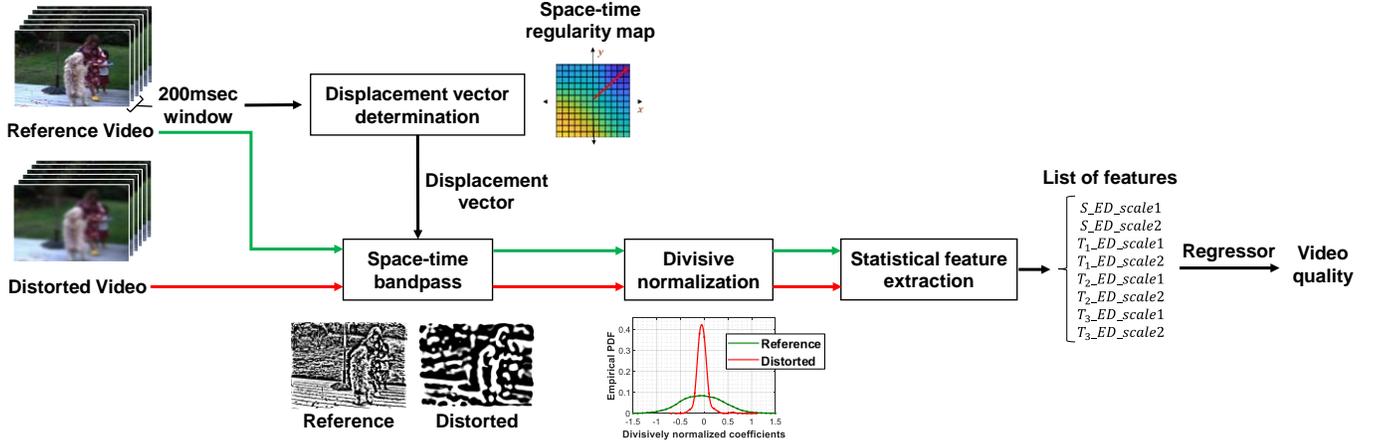

Fig. 4. Flowchart of the proposed quality model based on measurements of local spatial and temporal statistical irregularities.

Once the optimal displacement vectors are collected from all of the patches, construct a polar histogram to determine the dominant angle amongst the collected vectors. We used a polar histogram containing 48 bins, where each bin subtends an angular range of 7.5°. The average of all the local vectors that fall within the dominant angle bin is taken to be the optimal displacement vector for the one-second segment. This per-frame displacement vector corresponds to the optimal spatial displacement that is applied between any frame pairs having a temporal separation of 1 (i.e., adjacent frames) within the current one second segment. The divisively normalized bandpass coefficients computed from these displaced frame differences will possess quality-aware statistical information about possible loss of regularity arising from local video distortions.

*B. Bandpassed Plane Generation*

After the optimal displacement vectors are determined, generate multiple space-time bandpass planes from both the reference and the distorted videos. Fig. 5 depicts the four bandpass planes that are generated on each progression of frames.

**Spatial bandpass:** It has been amply shown that the bandpassed planes of pristine images or video frames reliably reflect an underlying Gaussianity that is revealed by divisive normalization, and that quantifying how distortions modify this Gaussian characteristic can be effectively used to measure perceptual spatial degradations [41], [43], [53]. Measuring frame-wise statistical losses of regularity make it possible to probe and measure perceptual degradations caused by spatial artifacts, which is an important aspect of video quality. Thus, generate a spatially bandpass plane on every reference and distorted video frame, by applying the local "Mean Subtraction" (MS) filter,

$$I_{ms}(i,j,k) = I(i,j,k) - \mu(i,j,k), \quad (6)$$

where $I(i,j,k)$ is luminance at pixel location $(i,j)$ of the $k^{th}$ video frame, and $\mu(i,j,k)$ is computed as in (4).

**Spatio-temporal bandpass:** As described in Section III, differencing frame patches that are displaced in space and time tends to reduce correlations between them. The existence of displaced space-time dependencies may also relate to hypothetical visual information-gathering processes involving small eye movement [44], [45] followed by bandpass, sparsifying temporal lag filtering in LGN. Thus, generate three spatio-temporally bandpass planes, by computing space-time displaced frame differences of Type 1. First, denote the optimal per-frame displacement vector determined on the reference video by the optimizing algorithm as $v = (v_x, v_y)$. Then, define three frame difference separations denoted as $T_1, T_2$, and $T_3$ which we used 1, 3, and 5, respectively. The space-time displaced frame difference planes are then computed using (1), and the space-time displacement vectors for each temporal separation $T_i$ are determined as $d_{T_i} = (v_x \times T_i, v_y \times T_i, T_i)$, for $i = 1, 2, 3$. Applying divisive normalization on these directionally bandpass filtered planes will ostensibly reveal Gaussianity of the processed reference video, and deviations from Gaussianity on the test video, if it is locally distorted. Statistical deviations between the coefficients of the two videos will generally relate to temporal or spatio-temporal aspects of video distortion.

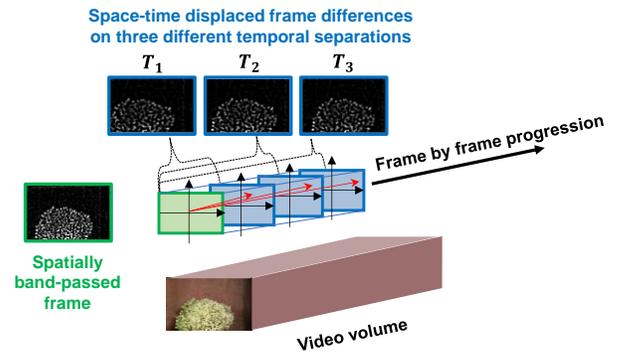

Fig. 5. Different types of bandpass planes generated on each progression of frames. One spatially bandpass plane is generated from local mean subtraction on the current frame. Three spatio-temporally bandpass planes are generated by computing space-time displaced frame differences using three different temporal separations ($T_1, T_2$, and $T_3$). The space-time displacement vector for each temporal separation is depicted by the red arrows.

## C. Statistical Feature Extraction

The bandpass, divisively normalized planes of the reference and distorted video coefficients are analyzed by computing entropic differences between the processed coefficients of the reference and distorted videos. These entropic differences are similar to those used in VIF [18], RRED [41], and ST-RRED [26] VQA models. Like these models, our approach relies on the Gaussian Scale Mixture (GSM) model of bandpass images used by these successful VQA models.

**GSM model of bandpass planes:** Many prior studies have shown that the bandpass coefficients of undistorted natural pictures, video frames, and frame differences reliably follow the Gaussian Scale Mixture (GSM) model [18], [26], [41], [54], [55]. Here, we expand the concept of GSM statistical regularity from spatial frames and frame differences, and posit that bandpass, space-time frame difference planes also contain space-time directional regularities accurately described by a GSM model. Let $p \in \{0, 1, 2, 3\}$ index the bandpass planes from the reference video, where $p$ corresponds to spatial bandpass and spatio-temporal bandpass planes at temporal separations $T_1, T_2$, and $T_3$, respectively. Partition the input bandpass planes into non-overlapping patches of size $\sqrt{N} \times \sqrt{N}$ (we take $N = 25$) indexed by $m \in \{1, 2, \ldots, M_p\}$. If we denote coefficients from the $m^{th}$ patch in bandpass plane $p$, generated from the $t^{th}$ frame of the reference video $(R)$ as $B_{mpt}^R$, then the coefficients within each patch may be modeled as

$$B_{mpt}^R = S_{mpt}^R U_{mpt}^R, \quad (7)$$

where $S_{mpt}^R$ is a scalar pre-multiplier random variable that is independent of the random field $U_{mpt}^R$, which is distributed as $U_{mpt}^R \sim \mathcal{N}(0, \mathbf{K}_{pt}^R)$ with covariance matrix $\mathbf{K}_{pt}^R$. Given a realization of the scalar $S_{mpt}^R = s_{mpt}^R$, then the distribution of the $m^{th}$ patch may be modeled as $B_{mpt}^R \sim \mathcal{N}(0, (s_{mpt}^R)^2 \mathbf{K}_{pt}^R)$. If we normalize the coefficients of each patch by the respective $s_{mpt}^R$, then $N_{mpt}^R = \frac{B_{mpt}^R}{s_{mpt}^R} \sim \mathcal{N}(0, \mathbf{K}_{pt}^R)$. Then, aggregating the divisively normalized coefficients $N_{mpt}^R$ over all patches within each bandpass plane, we expect the coefficients from each plane to follow a Gaussian distribution, corresponding to a spatial or space-time directional regularity observed on the reference video. Since $s_{mpt}^R$ is not known *a priori*, it must be estimated, which can be optimally accomplished via the Maximum Likelihood (ML) procedure:

$$\hat{s}_{mpt}^R = \text{argmax}_{(s_{mpt}^R)} p(B_{mpt}^R | S_{mpt}^R)$$
$$= \sqrt{\frac{(B_{mpt}^R)^T (\mathbf{K}_{pt}^R)^{-1} (B_{mpt}^R)}{N}}, \quad (8)$$

where $N$ is the number of coefficients within each patch, and $\hat{s}_{mpt}^R$ is the estimated normalization factor of the $m^{th}$ patch of the $p^{th}$ bandpass plane generated from the $t^{th}$ frame of the reference video. Similarly, we can model the bandpass coefficients of the distorted video $(D)$ as

$$B_{mpt}^D = S_{mpt}^D U_{mpt}^D, \quad (9)$$

and estimate the divisive normalization factor $\hat{s}_{mpt}^D$ as

$$\hat{s}_{mpt}^D = \text{argmax}_{(s_{mpt}^D)} p(B_{mpt}^D | S_{mpt}^D)$$
$$= \sqrt{\frac{(B_{mpt}^D)^T (\mathbf{K}_{pt}^D)^{-1} (B_{mpt}^D)}{N}}. \quad (10)$$

If distortion is present, then the bandpass planes of the distorted video may not follow a GSM distribution. Thus, GSM modeling of the bandpass planes of a distorted video may be considered as projecting the distorted video onto the space of natural undistorted videos. Distortions cause deviations from the regularity inherent in undistorted videos, which may be measured by a meaningful distance from the projection of the reference video. Since distortion may be viewed as visual information loss, following the VIF paradigm [18], we quantify the loss using entropic differencing.

**Entropic differencing:** We account for the uncertainties introduced on the observed reference (R) and distorted (D) videos by perceptual imperfections, such as neural noise along the visual pathway, by modeling the bandpass patches as passing through an additive Gaussian noise channel [18], [40],

$$\tilde{B}_{mpt}^R = B_{mpt}^R + W_{mpt}^R \text{ and } \tilde{B}_{mpt}^D = B_{mpt}^D + W_{mpt}^D, \quad (11)$$

where $W_{mpt}^R \sim \mathcal{N}(0, \sigma_W^2 \mathbf{I}_N)$, $W_{mi}^D \sim \mathcal{N}(0, \sigma_W^2 \mathbf{I}_N)$, $B_{mpt}^R$ is independent of $W_{mpt}^R$, $B_{mpt}^D$ is independent of $W_{mpt}^D$, and $W_{mpt}^R$ and $W_{mpt}^D$ are mutually independent. We fixed the neural noise variance at $\sigma_W^2 = 0.1$ as in [26], [27]. Let the eigenvalues of $\mathbf{K}_{pt}^R$ be $\beta_{1pt}^R, \beta_{2pt}^R, \ldots, \beta_{Npt}^R$, and those of $\mathbf{K}_{pt}^D$ be $\beta_{1pt}^D, \beta_{2pt}^D, \ldots, \beta_{Npt}^D$. Then, the local entropies $h$ of the data in the $m^{th}$ patch of the $p^{th}$ bandpass plane of the $t^{th}$ frame of the reference video are computed as follows:

$$h(\tilde{B}_{mpt}^R | S_{mpt}^R = \hat{s}_{mpt}^R)$$
$$= \frac{1}{2} \log[(2\pi e)^N |(\hat{s}_{mpt}^R)^2 \mathbf{K}_{pt}^R + \sigma_W^2 \mathbf{I}_N|]$$
$$= \sum_{n=1}^{N} \frac{1}{2} \log\left[(2\pi e)\left((\hat{s}_{mpt}^R)^2 \beta_{npt}^R\right) + \sigma_W^2\right], \quad (12)$$

and similarly, for the distorted video

$$h(\tilde{B}_{mpt}^D | S_{mpt}^D = \hat{s}_{mpt}^D)$$
$$= \frac{1}{2} \log[(2\pi e)^N |(\hat{s}_{mpt}^D)^2 \mathbf{K}_{pt}^D + \sigma_W^2 \mathbf{I}_N|]$$
$$= \sum_{n=1}^{N} \frac{1}{2} \log\left[(2\pi e)\left((\hat{s}_{mpt}^D)^2 \beta_{npt}^D\right) + \sigma_W^2\right]. \quad (13)$$

The entropies (12) and (13) are scaled by factors $\gamma_{mpt}^R = \log(1 + \hat{s}_{mpt}^{R\,2})$ and $\gamma_{mpt}^D = \log(1 + \hat{s}_{mpt}^{D\,2})$, respectively, which provides increased locality and emphasis on higher energy regions of the video [26], [27], [41]. The final entropic differences between the reference and distorted bandpass planes is given by:

$$ED_p = \frac{1}{M_p(T-T_3)} \sum_{m=1}^{M_p} \sum_{t=1}^{T-T_3} |\alpha_{mpt}^R - \alpha_{mpt}^D|, \quad (14)$$

where

$$\alpha_{mpt}^R = \gamma_{mpt}^R h(\tilde{B}_{mpt}^R | S_{mpt}^R = \hat{s}_{mpt}^R),$$
$$\alpha_{mpt}^D = \gamma_{mpt}^D h(\tilde{B}_{mpt}^D | S_{mpt}^D = \hat{s}_{mpt}^D),$$

TABLE I
SUMMARIZATION OF SPACE-TIME SUBSAMPLED VIDEO QUALITY DATABASES: AVT-VQDB-UHD-1 [15] AND ETRI-LIVE STSVQ [16]

| | Database attribute | AVT-VQDB-UHD-1 (test #4) [15] | ETRI-LIVE STSVQ [16] |
|---|---|---|---|
| Source information | Number of videos | 5 | 15 |
| | Resolution | 3840×2160 | 3840×2160 |
| | Frame rate | 60 Hz | 60/120 Hz |
| | Bit depth | 10 | 10 |
| | Chroma format | YUV422p | YUV 420p |
| | Average video length | 8 sec | 5.61 sec |
| Distortion information | Number of videos | 120 | 437 |
| | Spatial subsampling | 2160p → 1440/1080/720/480/360p | 2160p → 1080/720/540p |
| | Temporal subsampling | 60 Hz → 30/24/15 Hz | 60/120 Hz → 30/60 Hz |
| | Compression | HEVC (x265) compression to meet eight pre-defined target bit-rates (200, 500, 1000, 2000, 4000, 6000, 8000, 15000 kbps) | HEVC (x265) compression to meet five target bit-rate adaptively chosen per content to cover a wide range of qualities. |
| | Space-time resolution restoration | Spatial: Bicubic interpolation Temporal: Frame duplication | Spatial: Lanczos interpolation Temporal: Linear frame interpolation |
| Human study information | Protocol | Absolute Category Rating (ACR) | Single-Stimulus Continuous Quality Evaluation (SSCQE) with hidden reference |
| | Number of ratings | 3,000 (25 votes per video) | 13,560 (30 votes per video) |

TABLE II
CROSS-VALIDATION PERFORMANCE COMPARISON OF MODELS ON THE ETRI-LIVE STSVQ DATABASE ACROSS DIFFERENT TEMPORAL SUBSAMPLING LEVELS. THE NUMBERS DENOTE MEDIAN VALUES OVER 1000 ITERATIONS OF RANDOMLY SPLIT TRAIN AND TEST SETS. THE VALUES INSIDE THE PARENTHESES DENOTE STANDARD DEVIATIONS. THE TWO BEST MODELS IN EACH COLUMN ARE BOLDFACED.

| Model | Full frame rate | | Half frame rate | | Overall (full + half frame rate) | |
|---|---|---|---|---|---|---|
| | SRCC | PLCC | SRCC | PLCC | SRCC | PLCC |
| PSNR | 0.6805 (0.1506) | 0.6507 (0.1460) | 0.4930 (0.1806) | 0.3753 (0.2137) | 0.5092 (0.1282) | 0.4680 (0.1477) |
| SSIM [7] | 0.8136 (0.0871) | 0.7646 (0.1219) | 0.4623 (0.1988) | 0.2477 (0.2198) | 0.5317 (0.1391) | 0.3106 (0.2243) |
| MSSSIM [17] | 0.7492 (0.1370) | 0.7126 (0.1407) | 0.4654 (0.1841) | 0.2841 (0.2107) | 0.5186 (0.1299) | 0.3773 (0.1924) |
| VIF [18] | 0.7576 (0.1185) | 0.7335 (0.1365) | 0.5380 (0.1856) | 0.4481 (0.2485) | 0.5976 (0.1350) | 0.5399 (0.1774) |
| ST-RRED [26] | 0.8307 (0.1076) | 0.7344 (0.1506) | 0.4685 (0.1949) | 0.2510 (0.2221) | 0.5181 (0.1397) | 0.2615 (0.2008) |
| SpEED [27] | **0.8671 (0.0757)** | **0.7672 (0.1252)** | 0.4261 (0.1922) | 0.2380 (0.1886) | 0.4791 (0.1229) | 0.2079 (0.1670) |
| FRQM [32] | - | - | 0.1715 (0.1274) | 0.2089 (0.1391) | 0.1715 (0.1274) | 0.2089 (0.1391) |
| VMAF [8] | 0.7366 (0.1643) | 0.7353 (0.1614) | **0.6095 (0.2059)** | **0.6157 (0.2235)** | **0.6552 (0.1733)** | **0.6590 (0.1770)** |
| VSTR | **0.8876 (0.0705)** | **0.8871 (0.0810)** | **0.6401 (0.2004)** | **0.6646 (0.2249)** | **0.7702 (0.1137)** | **0.7767 (0.1210)** |

and $T$ refers to the total number of considered frames. The number of frames might be, for example, those from a single scene clip in a streaming scenario. Note that the last frame on which entropic differencing can be applied occurs at $T - T_3$, ensuring that all considered frames generate space-time displaced frame difference planes having temporal separation $T_3$. The values (14) computed for $p = 0, 1, 2$ and 3 each quantify how much spatial and space-time regularity at temporal separations of $T_1, T_2$, and $T_3$ are affected by distortion.

### D. Final Set of Features

In order to allow for the natural multi-scale behavior of both videos and distortions, as well as for variations of viewing conditions, we also applied the algorithm just described over multiple spatial resolutions [17]. Several previous studies have shown that features extracted at coarser scales generally outperform prediction power of the features extracted from a full resolution video [26], [27]. This may relate to the motion down-shifting phenomena [56], whereby the vision system becomes more sensitive lower spatial frequencies in the presence of motion, which are better represented at coarser scales. Similar to [26], [27], our model yielded higher performances at scales $k = 4$ and 5, where the vertical and horizontal dimensions of the video were each down-sampled by a factor of $2^k$, on which the entire feature extraction algorithm was applied. As explained in Section IV-C, since each pair of reference and distorted video results in four entropic difference values, operating at two scales yields a feature vector composed of eight elements. These features are combined using an SVR which learns to predict the final video quality. The nomenclature used for the final set of features follows the form of (Bandpass plane type)_ED_(scale type), where

- Bandpass plane type: the spatial bandpass case ($p = 0$) is denoted by '$S$,' while space-time displaced frame differences at varying temporal separations ($p = 1, 2,$ and 3) are each denoted by '$T_1$,' '$T_2$,' and '$T_3$.'
- Scale type: the case of scale factor $k = 4$ is denoted by 'scale 1,' while the case of scale factor $k = 5$ is denoted by 'scale 2.'

## V. EXPERIMENTAL RESULTS

Now we present and compare the prediction performance of our proposed features and the final video quality model (VSTR) against other leading video quality models. The models are evaluated on video quality databases having subjective quality scores rendered on videos subjected to spatial and/or temporal subsampling combined with compression.

### A. Experiment Setting

The video quality datasets that we used are AVT-VQDB-UHD-1 [15] and ETRI-LIVE STSVQ [16]. Table I summarizes the attributes of each database. Both databases first determine multiple target bit-rates to accommodate various bandwidth conditions. The designers of the AVT-VQDB-UHD-1 database used a universal set of eight target bit-rates identically applied on all source contents. The creators of the ETRI-LIVE STSVQ used a set of five target bit-rates adaptively chosen for each source content to cover a wide range of qualities, while also

9TABLE III
CROSS-VALIDATION PERFORMANCE COMPARISON OF MODELS ON THE ETRI-LIVE STSVQ DATABASE ACROSS DIFFERENT SPATIAL SUBSAMPLING LEVELS. THE
NUMBERS DENOTE MEDIAN VALUES OVER 1000 ITERATIONS OF RANDOMLY SPLIT TRAIN AND TEST SETS. THE VALUES INSIDE THE PARENTHESES DENOTE
STANDARD DEVIATIONS. THE TWO BEST MODELS IN EACH COLUMN ARE BOLDFACED.

|  | 540p | | 720p | | 1080p | | 2160p | | Overall | |
| --- | --- | --- | --- | --- | --- | --- | --- | --- | --- | --- |
|  | SRCC | PLCC | SRCC | PLCC | SRCC | PLCC | SRCC | PLCC | SRCC | PLCC |
| PSNR | 0.49 (0.17) | 0.45 (0.19) | 0.46 (0.17) | 0.42 (0.20) | 0.46 (0.16) | 0.40 (0.18) | 0.59 (0.13) | 0.53 (0.14) | 0.51 (0.13) | 0.47 (0.15) |
| SSIM [7] | 0.54 (0.18) | 0.34 (0.23) | 0.50 (0.16) | 0.26 (0.23) | 0.50 (0.18) | 0.24 (0.23) | 0.63 (0.12) | 0.40 (0.21) | 0.53 (0.14) | 0.31 (0.22) |
| MSSSIM [17] | 0.51 (0.17) | 0.39 (0.20) | 0.47 (0.17) | 0.34 (0.22) | 0.46 (0.16) | 0.30 (0.20) | 0.60 (0.12) | 0.43 (0.18) | 0.52 (0.13) | 0.38 (0.19) |
| VIF [18] | **0.63 (0.18)** | **0.59 (0.20)** | 0.59 (0.17) | 0.52 (0.22) | 0.56 (0.17) | 0.47 (0.21) | 0.68 (0.12) | 0.58 (0.16) | 0.60 (0.13) | 0.54 (0.18) |
| ST-RRED [26] | 0.51 (0.18) | 0.30 (0.19) | 0.47 (0.16) | 0.23 (0.20) | 0.49 (0.17) | 0.21 (0.20) | 0.62 (0.12) | 0.34 (0.22) | 0.52 (0.14) | 0.26 (0.20) |
| SpEED [27] | 0.45 (0.17) | 0.24 (0.17) | 0.42 (0.15) | 0.17 (0.16) | 0.44 (0.16) | 0.17 (0.16) | 0.60 (0.09) | 0.26 (0.17) | 0.48 (0.12) | 0.21 (0.17) |
| FRQM [32] | 0.26 (0.19) | 0.27 (0.20) | 0.20 (0.16) | 0.27 (0.17) | 0.20 (0.14) | 0.24 (0.17) | 0.12 (0.10) | 0.16 (0.11) | 0.17 (0.13) | 0.21 (0.14) |
| VMAF [8] | 0.54 (0.23) | 0.56 (0.25) | **0.63 (0.21)** | **0.63 (0.22)** | **0.64 (0.18)** | **0.62 (0.19)** | **0.75 (0.15)** | **0.74 (0.15)** | **0.66 (0.17)** | **0.66 (0.18)** |
| VSTR | **0.72 (0.16)** | **0.75 (0.16)** | **0.77 (0.15)** | **0.77 (0.15)** | **0.75 (0.13)** | **0.74 (0.14)** | **0.85 (0.10)** | **0.84 (0.10)** | **0.77 (0.11)** | **0.78 (0.12)** |

TABLE IV
RESULT OF WILCOXON RANKSUM TEST ON THE ETRI-LIVE STSVQ DATABASE. THE RESULTS ARE COMPUTED ON THE SRCC VALUES OF THE COMPARED MODELS
AT THE 95% CONFIDENCE LEVEL. EACH CELL CONTAINS 7 ENTRIES CORRESPONDING TO HALF FRAME RATE, FULL FRAME RATE, 540P, 720P, 1080P, 2160P AND ALL
VIDEOS. A SYMBOL '-' INDICATES STATISTICAL EQUIVALENCE BETWEEN THE ROW AND THE COLUMN. A VALUE '1' INDICATES THAT THE ROW MODEL WAS
STATISTICALLY SUPERIOR (BETTER QUALITY PREDICTION) THAN THE COLUMN MODEL. A VALUE '0' INDICATES THAT THE COLUMN MODEL WAS STATISTICALLY
SUPERIOR THAN THE ROW MODEL. A SYMBOL 'X' INDICATES CASE WHERE COMPARISON WAS IMPOSSIBLE, SINCE FRQM CANNOT BE COMPUTED AT FULL FRAME
RATES.

|  | PSNR | SSIM | MSSSIM | VIF | ST-RRED | SpEED | FRQM | VMAF | VSTR |
| --- | --- | --- | --- | --- | --- | --- | --- | --- | --- |
| PSNR | ------- | 1000000 | 100-000 | 0000000 | 100-000 | 1011-0- | 1X11111 | 0000000 | 0000000 |
| SSIM | 0111111 | ------- | -111111 | 0100000 | -011--- | 1011111 | 1X11111 | 0100000 | 0000000 |
| MSSSIM | 011-111 | -000000 | ------- | 0000000 | -0--001 | 10111-1 | 1X11111 | 0000000 | 0000000 |
| VIF | 1111111 | 1011111 | 1111111 | ------- | 1011111 | 1011111 | 1X11111 | 0-10000 | 0000000 |
| ST-RRED | 011-111 | -100--- | -1--110 | 0100000 | ------- | 1011111 | 1X11111 | 0100000 | 0000000 |
| SpEED | 0100-1- | 0100000 | 01000-0 | 0100000 | 0100000 | ------- | 1X11111 | 0100000 | 0000000 |
| FRQM | 0X00000 | 0X00000 | 0X00000 | 0X00000 | 0X00000 | 0X00000 | -X----- | 0X00000 | 0X00000 |
| VMAF | 1111111 | 1011111 | 1111111 | 1-01111 | 1011111 | 1011111 | 1X11111 | ------- | -000000 |
| VSTR | 1111111 | 1111111 | 1111111 | 1111111 | 1111111 | 1111111 | 1X11111 | -111111 | ------- |

ensuring a noticeable perceptual separation between the target bit-rates. The videos were then subsampled in space and/or time at various levels, as specified in Table I, followed by application of HEVC (libx265) compression to meet one of the defined target bit-rates. One thing to note is that the generated distorted videos were up-sampled back to the original source content's space-time resolution before being viewed. This space-time resolution restoration procedure is not only required for viewing, but also enables computation of Reference quality models that require pristine and distorted videos having the same spatial resolution and frame-rate. In regards to this, the two databases took slightly different approaches. The videos in the ETRI-LIVE STSVQ database were upscaled via Lanczos interpolation and linear frame interpolation, while those in the AVT-VQDB-UHD-1 database used Bicubic interpolation and frame duplication to restore the spatial and temporal dimensions of each video, respectively. The databases also differ in terms of the subjective experimental protocols that were used. AVT-VQDB-UHD-1 used the Absolute Category Rating (ACR) [57], whereby each participant evaluated video quality on a discrete 5 category scale, collected and converted to Mean Opinion Scores (MOS). ETRI- LIVE STSVQ adopted a Single-Stimulus Continuous Quality Evaluation (SSCQE) with hidden reference protocol [57], where the test participants used a continuous scale score bar to evaluate the video quality. The scores of the distorted videos and their respective reference videos were collected to compute Difference Mean Opinion Scores (DMOS). Since some database differences exist, the subjective scores rendered on each database may portray slightly different tendencies.

The space-time resolution adaptation framework is closely related to Reference quality models, since it considers how various dimension reduction methods affect the quality of a reference video. We compared the performances of nine relevant popular Reference video quality models, including PSNR, SSIM [7], MSSSIM [17], VIF [18], ST-RRED [26], SpEED [27], FRQM [32], VMAF [8], and the new model, VSTR. VMAF and VSTR are learning based models that each combines 6 and 8 spatio-temporal features, respectively, to predict final video quality scores.

The prediction performances of the compared quality models were evaluated using the Spearman's rank order correlation coefficient (SRCC) and Pearson linear correlation coefficient (PLCC). SRCC measure ordinal correlations, while PLCC measures linear correlations between variables. Higher values are favorable for both SRCC and PLCC. Before computing PLCC, the predicted scores from the various quality models were linearized using logistic regression, following the procedure in [57].

### B. Prediction Performances

Since the comparison models include the learning-based models VMAF and VSTR, we report the cross-validation performances so that the learning-based models can be properly evaluated by training on each respective database. Since both databases contain of videos afflicted by various combinations of dimension reduction methods applied on the same source contents, we took particular care to separate the train and test sets 'content-wise.' This means that the videos from train and test sets do not share videos having the same source contents.

On the ETRI-LIVE STSVQ database, which has a total of 15 source contents, we used 5-fold cross validation, where the model parameters were trained on videos generated from 12 source contents, and the performance was tested on videos from the other 3 source contents. The VMAF and VSTR models were both trained using an SVR with a radial basis function (RBF) kernel, where the SVR-RBF parameters were determined using cross validation within the training set, as described in [58]. We ran 1000 train-test iterations, where the train and test sets were randomly split over each iteration, while abiding by the content-wise separation.

TABLE V
CROSS-VALIDATION PERFORMANCE COMPARISON OF MODELS ON THE AVT-VQDB-UHD-1 DATABASE. THE NUMBERS DENOTE MEDIAN VALUES OVER 1000 ITERATION OF RANDOMLY SPLIT TRAIN AND TEST SETS. THE VALUES INSIDE THE PARENTHESES DENOTE STANDARD DEVIATIONS. THE TWO BEST MODELS IN EACH COLUMN ARE BOLDFACED.

| Model | SRCC | PLCC |
|---|---|---|
| PSNR | 0.5990 (0.1770) | 0.6566 (0.1443) |
| SSIM [7] | 0.7168 (0.1802) | 0.6798 (0.1420) |
| MSSSIM [17] | 0.6734 (0.1902) | 0.6882 (0.1756) |
| VIF [18] | 0.7009 (0.1827) | 0.6917 (0.1771) |
| ST-RRED [26] | 0.7216 (0.0761) | 0.6727 (0.0914) |
| SpEED [27] | 0.7484 (0.0916) | 0.6354 (0.1032) |
| FRQM [32] | 0.3425 (0.1405) | 0.4675 (0.0711) |
| VMAF [8] | **0.8387 (0.1743)** | 0.7670 (0.1586) |
| VSTR | **0.8004 (0.1133)** | **0.8178 (0.1344)** |

TABLE VI
RESULT OF WILCOXON RANKSUM TEST ON THE AVT-VQDB-UHD-1 DATABASE. THE RESULTS ARE COMPUTED ON THE SRCC VALUES OF THE MODELS AT THE 95% CONFIDENCE LEVEL. A SYMBOL '-' INDICATES STATISTICAL EQUIVALENCE BETWEEN THE ROW AND THE COLUMN. A VALUE '1' INDICATES THAT THE ROW MODEL WAS STATISTICALLY SUPERIOR (BETTER QUALITY PREDICTION) THAN THE COLUMN MODEL. A VALUE '0' INDICATES THAT THE COLUMN MODEL WAS STATISTICALLY SUPERIOR THAN THE ROW MODEL.

|  | PSNR | SSIM | MSSSIM | VIF | ST-RRED | SpEED | FRQM | VMAF | VSTR |
|---|---|---|---|---|---|---|---|---|---|
| PSNR | - | 0 | 0 | 0 | 0 | 0 | 1 | 0 | 0 |
| SSIM | 1 | - | 1 | 1 | 0 | 0 | 1 | 0 | 0 |
| MSSSIM | 1 | - | - | 0 | 0 | 0 | 1 | 0 | 0 |
| VIF | 1 | 0 | 1 | - | - | - | 1 | 0 | 0 |
| ST-RRED | 1 | 1 | 1 | - | - | 0 | 1 | 0 | 0 |
| SpEED | 1 | 1 | 1 | - | 1 | - | 1 | 0 | 0 |
| FRQM | 0 | 0 | 0 | 0 | 0 | 0 | - | 0 | 0 |
| VMAF | 1 | 1 | 1 | 1 | 1 | 1 | 1 | - | - |
| VSTR | 1 | 1 | 1 | 1 | 1 | 1 | 1 | - | - |

Table II the reports median and standard deviations of prediction performance across the 1000 train-test splits over different temporal subsampling levels and overall. The median performances of other non-learning-based models are reported on the same randomized splits for comparison. Since FRQM requires the distorted videos to have lower frame rates than the reference video, we only report its performance for the half frame rate case. An interesting tendency observed is that the compared models attained relatively high performances on the full frame rate case as compared to the half frame rate case. The full frame rate performances were computed from the subset of test videos containing only spatial subsampling and compression as distortion types. As the results suggest, most of the models were able to accurately predict the quality of videos afflicted by mixtures of these two distortions. However, the performances of most models fell considerably when temporal subsampling was also introduced. This is likely because of introduced temporal effects such as stutter. This suggest that there is ample room for improvement of the models to predict the quality of temporally subsampled and compressed videos. Among the models, the learning-based methods maintained high prediction performances on both the half frame rate and overall cases. In particular, the proposed VSTR outperformed the other models by a wide margin.

TABLE VII
CROSS-VALIDATION PERFORMANCE COMPARISON OF MODELS ON THE MCL-V DATABASE. THE NUMBERS DENOTE MEDIAN VALUES OVER 1000 ITERATION OF RANDOMLY SPLIT TRAIN AND TEST SETS. THE VALUES INSIDE THE PARENTHESES DENOTE STANDARD DEVIATIONS. THE TWO BEST MODELS IN EACH COLUMN ARE BOLDFACED.

| Model | SRCC | PLCC |
|---|---|---|
| PSNR | 0.5363 (0.1046) | 0.5435 (0.1065) |
| SSIM [7] | 0.7185 (0.1108) | 0.7363 (0.1074) |
| MSSSIM [17] | 0.6653 (0.1088) | 0.6955 (0.1068) |
| VIF [18] | 0.7533 (0.0989) | 0.7506 (0.0981) |
| ST-RRED [26] | 0.8032 (0.1111) | 0.8080 (0.1075) |
| SpEED [27] | **0.8398 (0.0916)** | **0.8383 (0.0890)** |
| VMAF [8] | 0.8358 (0.0768) | 0.8245 (0.0778) |
| VSTR | **0.8515 (0.0704)** | **0.8539 (0.0748)** |

Table III reports cross-validation performance over different spatial subsampling levels and overall. Unlike the case for separation by different temporal subsampling levels, here we observe similar prediction performances across all resolutions. VIF delivered good performance at low spatial resolution (540p), while VMAF yielded good performances at higher spatial resolutions (720p, 1080p, and 2160p). Overall, VSTR delivered the best performances across all resolutions.

We verified the statistical significance of the performance differences among the compared models in Tables II and III, using the distribution of the SRCC scores computed on 1000 random train-test splits. Table IV shows the results of a Wilcoxon ranksum test [59] performed on the SROCC distributions of pairs of models. The null hypothesis was that the median of the row model and the column model were equal (or indistinguishable) at the 95% confidence level, which is indicated by a symbol '-' in the table. The alternate hypothesis states that the median of the row model and the column model were different at a statistically significant level, where a value '1' indicates the row model had higher median values as compared to the column model, while a value '0' indicates otherwise. Note that the symbol 'x' indicates cases where comparison was impossible, since FRQM cannot be computed on the full frame rate case. Table IV contains 7 entries per cell corresponding to half frame rate, full frame rate, 540p, 720p, 1080p, 2160p, and all of the videos, in that order. As shown in the Table, the proposed VSTR model attained statistically superior prediction performance as compared to the other models.

The AVT-VQDB-UHD-1 database has a total of 5 source contents. We used a 3-to-2 train-test split, where the model parameters were trained on videos from the 3 source contents and the tested on the videos from the other 2 source contents. On average, we tested 48 distorted videos per iteration, which were generated from 2 source contents. We only reported overall performances, since grouping the test videos along different space or time subsampling levels would result in correlations being computed on too small a number of data points (<10). Table V and VI show the cross-validation performances over 1000 random train-test splits and the statistical significance of the performance differences among the compared models. The overall performances of all models appeared higher as compared to the results on ETRI-LIVE STSVQ, which may be caused by different contents, human study protocol, space-time interpolation methods, and train-test split ratio. Regardless, we still observe similar model tendencies whereby VMAF and VSTR yielded statistically superior prediction performances as compared to the other models.

TABLE VIII
CROSS-VALIDATION PERFORMANCE COMPARISON OF MODELS ON THE LIVE-YT-HFR DATABASE ACROSS DIFFERENT FRAME RATES. THE NUMBERS DENOTE MEDIAN VALUES OVER 1000 ITERATIONS OF RANDOMLY SPLIT TRAIN AND TEST SETS. THE VALUES INSIDE THE PARENTHESES DENOTE STANDARD DEVIATIONS. THE TWO BEST MODELS IN EACH COLUMN ARE BOLDFACED.

| | 24 fps | | 30 fps | | 60 fps | | 82 fps | | 98 fps | | 120 fps | | Overall | |
|---|---|---|---|---|---|---|---|---|---|---|---|---|---|---|
| | SRCC | PLCC | SRCC | PLCC | SRCC | PLCC | SRCC | PLCC | SRCC | PLCC | SRCC | PLCC | SRCC | PLCC |
| PSNR | 0.57 (0.19) | 0.49 (0.16) | 0.55 (0.19) | 0.52 (0.14) | 0.64 (0.15) | 0.63 (0.12) | 0.70 (0.13) | 0.69 (0.11) | 0.73 (0.13) | 0.69 (0.13) | 0.75 (0.15) | 0.73 (0.14) | 0.76 (0.08) | 0.73 (0.07) |
| SSIM [7] | 0.47 (0.22) | 0.29 (0.15) | 0.47 (0.22) | 0.28 (0.14) | 0.49 (0.19) | 0.33 (0.16) | 0.53 (0.17) | 0.44 (0.19) | 0.65 (0.17) | 0.60 (0.19) | 0.82 (0.16) | 0.80 (0.19) | 0.64 (0.09) | 0.53 (0.12) |
| MSSSIM [17] | 0.48 (0.20) | 0.33 (0.16) | 0.45 (0.21) | 0.29 (0.14) | 0.45 (0.21) | 0.35 (0.16) | 0.51 (0.18) | 0.43 (0.18) | 0.58 (0.18) | 0.57 (0.18) | 0.73 (0.13) | 0.72 (0.12) | 0.64 (0.09) | 0.57 (0.11) |
| VIF [18] | 0.51 (0.21) | 0.43 (0.16) | 0.49 (0.22) | 0.45 (0.15) | 0.60 (0.22) | 0.58 (0.17) | 0.70 (0.18) | **0.74 (0.14)** | **0.83 (0.15)** | **0.87 (0.11)** | **0.82 (0.10)** | **0.85 (0.08)** | 0.75 (0.08) | 0.71 (0.08) |
| ST-RRED [26] | 0.38 (0.20) | 0.28 (0.16) | 0.29 (0.20) | 0.18 (0.14) | 0.49 (0.16) | 0.44 (0.19) | 0.50 (0.21) | 0.50 (0.19) | 0.55 (0.20) | 0.58 (0.20) | 0.74 (0.15) | 0.65 (0.15) | 0.61 (0.07) | 0.52 (0.09) |
| SpEED [27] | 0.39 (0.20) | 0.25 (0.15) | 0.31 (0.21) | 0.20 (0.15) | 0.28 (0.22) | 0.22 (0.18) | 0.40 (0.22) | 0.35 (0.21) | 0.46 (0.22) | 0.42 (0.22) | 0.74 (0.15) | 0.65 (0.15) | 0.56 (0.08) | 0.49 (0.09) |
| FRQM [32] | 0.26 (0.16) | 0.17 (0.13) | 0.27 (0.16) | 0.19 (0.11) | 0.29 (0.17) | 0.22 (0.15) | 0.38 (0.18) | 0.24 (0.15) | 0.48 (0.19) | 0.32 (0.17) | - | - | 0.57 (0.09) | 0.54 (0.09) |
| VMAF [8] | **0.62 (0.22)** | **0.62 (0.21)** | **0.61 (0.24)** | **0.62 (0.18)** | 0.66 (0.22) | **0.73 (0.17)** | **0.75 (0.16)** | **0.79 (0.12)** | 0.76 (0.13) | **0.78 (0.11)** | 0.75 (0.12) | 0.78 (0.11) | **0.81 (0.11)** | **0.78 (0.09)** |
| VSTR | **0.63 (0.22)** | **0.57 (0.19)** | **0.59 (0.24)** | **0.59 (0.20)** | **0.67 (0.23)** | **0.71 (0.19)** | **0.72 (0.17)** | 0.71 (0.14) | **0.80 (0.13)** | 0.77 (0.11) | **0.82 (0.11)** | **0.84 (0.10)** | **0.78 (0.12)** | **0.76 (0.11)** |

*C. Performances on Other Space/Time VQA Databases*

We investigated the generalizability of VSTR by evaluating its prediction performances on two other VQA databases: MCL-V [9] and LIVE-YT-HFR [33]. These databases consider different combinations of distortion types, where each focus on how one kind of subsampling, spatial or temporal, affects perceptual quality when combined with compression.

The MCL-V database has 12 source contents of resolution 1920×1080 and frame rates 24~30 fps. A total of 96 distorted videos were generated from these source contents by applying spatial subsampling and AVC (x264) compression. We used an 8-to-4 train-test split, where the model parameters were trained on videos from the 8 source contents and tested on the videos from the other 4 source contents. Table VII shows the cross-validation performances over 1000 random train-test splits. From the Table, it may be seen that VSTR and SpEED yielded superior prediction performances as compared to the other models.

The LIVE-YT-HFR database has 16 source contents of resolutions 3840×2160 or 1920×1080 and frame rate 120 fps. A total of 480 distorted videos were generated by jointly applying temporal downsampling and VP9 compression. Table VIII reports the cross-validation performances across all frame rates and overall, over 1000 random train-test splits. We used 12-to-4 train-test splits. From the Table, it may be observed that VSTR yielded competitive performances compared to the other models. The learning-based models, VSTR and VMAF delivered high performance across frame rates, indicating robustness to frame rate variations.

## VI. CONCLUSION

We proposed a new video quality model called VSTR, that can account for the perceptual effects of space-time distortions such as spatial and/or temporal subsampling and compression applied in concert. VSTR is based on new findings on the space-time statistics of natural videos, where we showed the existence of space-time directional regularities which are revealed by differencing frames that are displaced in space and time, followed by divisive normalization.

Motivated by these findings, we devised a way to identify optimal space-time regular paths of a pristine video. We derived features that quantify how these space-time directional regularities are disturbed by space-time distortions such as spatial and/or temporal subsampling and compression. These statistical features are combined using an SVR to produce a video quality prediction model called VSTR.

Comprehensive performance evaluations against human subjective scores drawn from relevant video quality databases show that VSTR is able to deliver accurate predictions on videos afflicted by various levels of space-time subsampling and compression. VSTR was able to provide robust prediction performance across a variety of spatial and temporal subsampling levels, and outperformed other models by a wide margin. We envision that VSTR can be used to assist optimal space-time resolution adaptation strategies for perceptual video compression.

ACKNOWLEDGMENT

This work was supported by an Institute for Information & Communications Technology Promotion (IITP) grant funded by the Korean government (MSIT) (No. 2017-0-00072, Development of Audio/Video Coding and Light Field Media Fundamental Technologies for Ultra Realistic Teramedia).